\documentclass[a4paper,11pt]{article}
\usepackage{pos}

\title{Application of the projective truncation and randomized singular value decomposition to a higher dimension}

\author*[a]{Katsumasa Nakayama}

\affiliation[a]{RIKEN Center for Computational Science, Kobe 650-0047, Japan}

\emailAdd{katsumasa.nakayama@riken.jp}

\abstract{We study the tensor renormalization group (TRG) in the dimension larger than two as the Higher-order TRG (HOTRG) with the randomized SVD method. The randomized SVD and the detailed discussion on the low order tensor representation, we can calculate the HOTRG with the reduced computational cost. We also represent our method by using the cost function, and the details of the cost function for the isometry determine the precision, stability, and calculation time. In our study, we show calculation order improvement using randomized SVD. We also propose that the internal line respect for any TRG method improves the calculation without changing the order of the computational cost.}

\FullConference{The 40th International Symposium on Lattice Field Theory (Lattice 2023)\\
July 31st - August 4th, 2023\\
Fermi National Accelerator Laboratory\\}

\begin{document}
\maketitle

\section{Introduction}
The tensor renormalization group (TRG) method is a method to calculate the physical quantities by numerical calculation \cite{TRG}.
Because the TRG method is free from the sign problem, we can apply the TRG as a good candidate to produce physical quantities in the sign-problematic system.
There has been related progress in formulation and application in the past decade.
Especially in the higher dimensional system, HOTRG is formulated by the paper \cite{HOTRG} and applied to the three or four-dimensional system to calculate physical quantities such as the partition function.

Although the HOTRG method is the most straightforward application of the TRG to the higher dimensional system, it still suffers from the large computational cost.
In this proceeding, we mention that the cut-off bond size is $D$. 
The HOTRG requires the computational cost $O(D^{4d - 1})$ in a d-dimensional system, and then it is difficult to set the bond size $D$ large enough to control the systematic error from the truncation.

In order to reduce the computational cost in the higher dimension, more sophisticated algorithms are developed, such as the anisotropic TRG (ATRG) and the Triad TRG(TTRG) \cite{ATRG, TriadTRG}.
ATRG and TTRG introduce additional decomposition to represent tensor networks by lower-order tensors, which can reduce the cost of contraction and decomposition.
These methods can reduce the order of the cost as $O(D^{2d+1})$ for ATRG and $O(D^{d+3})$ for TTRG.

The additional decomposition achieves the cost reduction. However, it becomes an additional resource for the systematic error in the calculation.
ATRG and TTRG clearly show additional systematic errors in the three-dimensional Ising model in the original papers.
Because the trade-off between the cost reduction and the additional systematic error is not well understood, we need to do practical calculations for each system to confirm that our calculations are more reliable than that of HOTRG.
We still investigate the precise understanding of the systematic error and try to find a new formulation of the TRG in the higher dimensions.

In this proceedings, we apply randomized SVD \cite{RSVD} to the HOTRG to reduce the computational cost. The cost is reduced from $O(D^{4d - 1})$ to $O(D^{3d})$, although the dominant part of the HOTRG computational cost comes from the contraction step.
Randomized SVD allows for the reduction of the contraction cost, and then our randomized HOTRG method achieves the cost reduction.
We also introduce further cost reduction considering the lower-order tensor representation of the randomized HOTRG method.
We can find the lower order tensor representation of the randomized HOTRG without additional decomposition, and we call it a minimally decomposed TRG method (MDTRG) with the cost $O(D^{2d + 1})$.
In the MDTRG method, we also need a precise discussion of the oversampling in the step of the randomized SVD.

In order to further reduce costs, we introduce additional decomposition concerning the original approximation region of HOTRG.
This additional decomposition produces order three (triad) tensor representation, and then this Triad-MDTRG method requires the $O(D^{d + 3})$ cost.
In the Triad-MDTRG method, we also introduce the cost function representation of the Isometry and consider the approximation tensor network region.
The cost function representation produces a clear understanding of the approximation region of the corresponding decomposition \cite{TNR, PTTRG}.
It is also helpful to consider the projective truncation in any TRG method.
This proceeding is based on the paper \cite{MDTRG}.

\section{Cost function representation}

We introduce cost function representation in TRG and HOTRG as the first step.
In the simple TRG method, we consider the partition function $Z$ in the tensor network representation as follows,
\begin{align}
Z &\equiv
\mathrm{Tr}
\sum_{i}A_{x_iy_ix' _i y' _i},
\label{eq:Tr}
\end{align}
where the indices $i$ correspond to each lattice point.
Note that we assume the tensor network has homogeneity, and then the tensor network can be represented by the $A$.

The TRG method is usually represented by the SVD and contraction.
We introduce the different SVD for even and odd lattice points,
\begin{align}
A_{x_iy_ix' _i y' _i}
&=
\sum_{k} ^{D^2}
U^{\mathrm{(odd)}} _{x_iy_ik}
\lambda_k ^{\mathrm{(odd)}}
V^{\mathrm{(odd)}} _{x' _iy' _ik},\nonumber\\
A_{x_iy_ix' _i y' _i}
&=
\sum_{k} ^{D^2}
U^{\mathrm{(even)}} _{x_iy' _ik}
\lambda_k ^{\mathrm{(even)}}
V^{\mathrm{(even)}} _{x' _iy_ik},
\label{eq:decodd}
\end{align}
where the $U$ and $V$ are unitary matrix and the $\lambda$ is a singular value of the decomposition.
In order to take the contraction of the tensor network approximately, we cut-off the sum up to $D$, and we construct the next tensor network by coarse-grained tensor $A^\mathrm{(next)}$,
\begin{align}
A_{XYX' Y' } ^\mathrm{(next)}
&=
\sum_{x_1,x_2,y_1,y_2}
U^{\mathrm{(odd)}} _{x_1y_1X}
U^{\mathrm{(even)}} _{x_2y _1Y}
V^{\mathrm{(odd)}} _{x _2y _2X'}
V^{\mathrm{(even)}} _{x _1y _2Y'}
\sqrt{\lambda_{X} ^{\mathrm{(odd)}}\lambda_{Y} ^{\mathrm{(even)}}\lambda_{X'} ^{\mathrm{(odd)}}\lambda_{Y'} ^{\mathrm{(even)}}}.
\label{eq:nextA}
\end{align}

This decomposition and contraction are also represented by isometry, which was first introduced to TRG as HOTRG, although it is a redundant representation for the simple TRG method.
The isometry $U$ and $V$ for equation (\ref{eq:decodd}) is nothing but the truncated unitary matrix $U^{\mathrm{(odd)}}$ and $V^{\mathrm{(odd)}}$ in odd lattice point.
We can write the corresponding cost function,
\begin{align}
|| 
UU^tAVV^t - A
||
\label{eq:TRGcf}
\end{align}
In this proceeding, we abbreviate the contraction of the indices, but we can easily identify the contracted indices from the definition of the isometries.
If we set $U$ and $V$ using truncated SVD, this procedure is the same for the simple TRG method.
In addition, we can optimize the isometry $U$ and $V$ by using variational optimization.
Variational optimization is widely applied in the tensor network with Hamiltonian formalisms \cite{MPSPEPS} and tensor network renormalization in Lagrangian formalism \cite{TNR}.
For the simple TRG method, it is also studied as the projective truncation method \cite{PTTRG}, and produces the same precision as the original simple TRG method.

We also consider the HOTRG in the cost function.
In the HOTRG, we introduce the approximated region called "unit-cell tensor network" $\Gamma$ in the method.
\begin{align}
\Gamma_{x_1x' _1y' _1x_2y _2x' _2}
=
\sum_{y}
A_{x_1yx' _1y' _1}
A_{x_2y_2x' _2y}.
\end{align}
We find the isometry $U=V$ for the HOTRG by using the unit-cell tensor $\Gamma$ as follows, 
\begin{align}
\sum_{a,b,c,d}
\Gamma_{x_1abx_2cd}
\Gamma_{x_1 ^tabx_2 ^tcd}
=
\sum_{k}
U^{\mathrm{(HOTRG)}} _{x_1x_2k}
\lambda_k ^{\mathrm{(HOTRG)}}
U^{\mathrm{(HOTRG)}} _{x _1 ^tx _2 ^tk}.
\end{align}
We construct the next tensor network by coarse-grained tensor $A^\mathrm{(next)}$,
\begin{align}
A_{Xy_2X' y' _1} ^\mathrm{(next)}
=
\sum_{x_1,x_2,x' _1,x' _2}
U^{\mathrm{(HOTRG)}} _{x_1x_2X}
U^{\mathrm{(HOTRG)}} _{x' _1x' _2X'}
\Gamma_{x_1x' _1y' _1x_2y _2x' _2}.
\end{align}
We also find the isometry $U$ for the HOTRG by using the unit-cell tensor $\Gamma$ as follows, 
\begin{align}
|| 
UU^t\Gamma - \Gamma
||.
\label{eq:cf}
\end{align}
This cost function representation for the SVD is a general form.
For example, if we consider the unit-cell tensor $\Gamma = A$ and $U_{xyXx'y'X'} = U^\mathrm{(odd)} _{xyX}V^\mathrm{(odd)} _{x'y'X'}$, the cost function (\ref{eq:cf}) is exactly same to the equation (\ref{eq:TRGcf}).

We also use this cost function representation to understand the approximation region.
In any decomposition and contraction, we consider the approximated region as the unit-cell tensor network $\Gamma$.
In this proceedings, although the representation becomes a lower-order tensor network compared to the original HOTRG $\Gamma = AA$, we consider the same approximated region $\Gamma = AA$.
This property is one typical difference from the other methods, such as the ATRG and TTRG.
In order to maintain the same approximated region $\Gamma$, we apply the randomized SVD method to the decomposition and contraction step.

\section{Randomized SVD}

Randomized SVD is a method to take truncated SVD with reduced calculation cost.
We use the random samplings and QR decomposition to find the isometry corresponding to the truncated SVD.
The details of this method are introduced in previous works, and these works also show that precision only depends on the truncated singular values \cite{RSVD, RandTRG}.
We also utilize the randomized SVD method to approximate the contraction in this proceeding.
This method is also introduced in the TTRG method; details are in the papers \cite{TriadTRG, MDTRG}.

The simplest SVD is the SVD of $D^2\times D^2$ matrix $A = U\lambda V$ as shown in the equation (\ref{eq:decodd}).
We consider the sample matrix $\Theta = A\Omega$ by using $D^2 \times D$ random matrix $\Omega$.
The QR decomposition of the sample matrix $\Theta = QR$ produces the truncated orthogonal matrix $Q$, which has the approximated orthogonality $QQ\dagger \simeq I$.
Using this truncated orthogonal matrix $Q$, we can approximate the original matrix $A \simeq QQ^\dagger A$.
Since the truncated orthogonal matrix, $Q$ is $D^2 \times D$ matrix, contraction of the $A$ and $Q^\dagger$ need only $O(D^5)$ cost.
After taking the contraction, we can perform the SVD of $Q^\dagger A = \tilde{U}\lambda V$ and define the isometry $U \simeq Q\tilde{U}$ to approximate the original SVD $A = U\lambda V$.
Comparing the order estimate of the original SVD cost $O(D^6)$, the computational cost of the randomized SVD is reduced.

We emphasize that the randomized SVD can be regarded as the approximation of the contraction, not only the approximation of the decomposition.
In the approximation step by equation $A \simeq QQ^\dagger A$, the matrix $A$ can be considered constructed by several matrices.
For example, we consider that $A$ is defined by the contraction of the different matrices $B$ and $C$ as $A = BC$, and then the equation becomes the approximation of the contraction of $B$ and $C$.
\begin{align}
BC \simeq QQ^\dagger BC
\end{align}
It also means that the corresponding cost function is $||QQ^\dagger A - A||$, and the $Q$ is nothing but the corresponding isometry.
This representation also could reduce the computational cost.
For example, for the $D^2 \times D^2$ matrices $B$ and $C$, the contraction $Q[(Q^\dagger B)C]$ can be performed only the $O(D^5)$ cost.

For the precision of the randomized SVD at $A \simeq QQ^\dagger A$, we have to require several conditions and detailed techniques, described in \cite{RSVD}.
First, the truncated singular values should be small enough compared to the dominant part.
This assumption is always required in any TRG method, and often, we mention it as the hierarchy condition of the SVD.
Because this hierarchy problem is always assumed in the TRG, this is not the special constraint of the randomized SVD.
Second, we have to introduce the oversampling parameter $r$ in the index size of the random matrix $\Omega$.
In order to respect the precision up to bond dimension $D$, we have to set the size of the random matrix $\Omega$ as $D^2 \times rD$.
We must choose the oversampling parameter $r$ to achieve the required precision. 
Third, we can improve the precision by using iterative samplings.
Once we get the truncated orthogonal matrix $Q$, we can resample the original matrix $A$ by using $Q$ as $\Theta ' = Q^\dagger A$. Again, we consider the QR decomposition of the $\Theta ' {}^\dagger= Q'R'$ to approximate the original matrix $A \simeq A Q'Q' {}^\dagger$.
We must also choose the optimal number of the QR decomposition $q$ as the iterative samplings.
In this proceedings, we set the oversampling parameter $r = 6$ and the number of the QR decomposition $q = 2$ unless otherwise mentioned.

\section{Randomized HOTRG}

In this section, we apply the randomized SVD to the contraction step of the HOTRG.
Because the dominant computational cost originated from the contraction step, the cost reduced from $O(D^{4d-1})$ to $O(D^{3d})$.

In the d-dimensional system, the unit-cell tensor becomes an order $2d$ tensor, which has the independent $2d$ indices with bond dimension $D$.
In the isometry step for x-direction, we need to calculate $[\Gamma \Gamma^\dagger]_{x_1x_2x' _1x' _2}$ by the contraction of other $2d-2$ indices. 
The corresponding computational cost is only $O(D^{2d+2})$, which is not a dominant part of the calculation.
In the contraction step, we have to calculate $U^\mathrm{(HOTRG) \dagger}\Gamma U^\mathrm{(HOTRG)} = U^\mathrm{(HOTRG) \dagger}AA U^\mathrm{(HOTRG)} = A^\mathrm{(next)}$.
The isometry $U^\mathrm{(HOTRG)}$ is defined by the isometries of $d-1$ in different directions.
For example, in a three-dimensional system, we calculate the isometries of the $x$ and $y$ direction as $U^{(x)} _{x_1x_2X}$ and $U^{(y)} _{y_1y_2Y}$ and define the total isometry $U^\mathrm{(HOTRG)} _{x_1x_2X y_1y_2Y}= U^{(x)} _{x_1x_2X}U^{(y)} _{y_1y_2Y}$.
The contraction cost of $AU^\mathrm{(HOTRG)}$ and $U^\mathrm{(HOTRG)\dagger}A$ is $O(D^{3d-1})$, and cost of $[U^\mathrm{(HOTRG) \dagger}A][A U^\mathrm{(HOTRG)}]$ is $O(D^{4d - 1})$.
The last step is the dominant part of the computational cost in the HOTRG method.

We can straightforwardly apply randomized SVD to the dominant part which is the matrix-matrix product of $D^{d} \times D^{2d-1}$ matrix $[U^\mathrm{(HOTRG) \dagger}A]$ and $D^{2d-1} \times D^{d}$ matrix $[A U^\mathrm{(HOTRG)}]$.
We substitute $B = [U^\mathrm{(HOTRG) \dagger}A]$ and $C = [A U^\mathrm{(HOTRG)}]$, and consider the random matrix $\Omega$ with the size $D^d \times rD$ to prepare the truncated orthogonal matrix $Q$ with the same size $D^d \times rD$.
The corresponding cost function for $Q$ is 
\begin{align}
||
QQ^\dagger U^\mathrm{(HOTRG) \dagger}\Gamma U^\mathrm{(HOTRG)}
-
U^\mathrm{(HOTRG) \dagger}\Gamma U^\mathrm{(HOTRG)}
||.
\end{align}
We can use the matrix $Q$ to take the approximate contraction with the computational cost $O(D^{3d})$.
It also means that the course-grained tensor is 
\begin{align}
A^\mathrm{(next)}
\simeq
QQ^\dagger U^\mathrm{(HOTRG) \dagger}\Gamma U^\mathrm{(HOTRG)}.
\label{RHOTRG}
\end{align}

\section{Minimally-decomposed TRG}

In order to realize further cost reduction, we have to reduce the order of the tensor in the contraction and decomposition.
The unit-cell tensor $\Gamma$ is contracted by order ${2d}$ tensor $A$, determining the HOTRG method's computational cost with and without randomized SVD.
The lower-order representation of the $A$ can reduce the computational cost, although the precision may be sacrificed depending on how to find the lower-order representation.
How to prepare the lower-order representation is crucial for the precision and cost in any TRG algorithm.
If we introduce additional decomposition, it could be an additional resource of the systematic error.

In the randomized HOTRG, however, we can automatically prepare the lower-order representation of the $A^\mathrm{(next)}$ as shown in the equation (\ref{RHOTRG}).
The tensor $Q$ and $\Lambda \equiv [Q^\dagger U^\mathrm{(HOTRG) \dagger}\Gamma U^\mathrm{(HOTRG)}]$ are order $d+1$, these are already lower-order representation of $A^\mathrm{(next)} = Q\Lambda$ without any additional decomposition.
Note that the tensor $Q$ and $\Lambda$ have the size $rD$ oversampled index, which is the internal line between the $Q$ and $\Lambda$. This internal line oversampling is essential to calculate with the same precision as the original HOTRG.

For the order $d+1$ representation, we define the $\Gamma ^{(EFGH)} \equiv EFGH \simeq \Gamma$ for order $d+1$ tensors $E = G = Q$ and $F = H = \Lambda$.
The total isometry using $\Gamma ^{(EFGH)}$ is defined as $U^\mathrm{(MDTRG)}$ with $O(r^2D^{d+3})$ cost.
For example, $U^\mathrm{(MDTRG)} _{x_1x_2Xy_1y_2Y}= U^\mathrm{'(x)} _{x_1x_2X}U^\mathrm{'(y)} _{y_1y_2Y}$ in the three dimensional system, where the isometries $U^\mathrm{'(x)}$ and $U^\mathrm{'(y)}$ for each direction are calculated by the $\Gamma ^{(EFGH)} \Gamma ^{(EFGH)\dagger}$.
The corresponding cost function for $Q$ is 
\begin{align}
||
QQ^\dagger U^\mathrm{(MDTRG) \dagger}\Gamma ^{(EFGH)} U^\mathrm{(MDTRG)}
-
U^\mathrm{(MDTRG) \dagger}\Gamma ^{(EFGH)} U^\mathrm{(MDTRG)}
||.
\end{align}
We can use the matrix $Q$ to take the approximate contraction with the computational cost $O(D^{3d})$.
It also means that the course-grained tensor is 
\begin{align}
A^\mathrm{(next)}
\simeq
QQ^\dagger U^\mathrm{(MDTRG) \dagger}\Gamma ^{(EFGH)} U^\mathrm{(MDTRG)}.
\label{MDTRG}
\end{align}
The computational cost of the contraction step is $O(r^2D^{2d+1})$, which is in the same order as the ATRG.
We call this method minimally decomposed TRG (MDTRG) since we do not introduce additional decomposition compared to the randomized HOTRG.

\begin{figure}[]
 \centering
 \includegraphics[clip,width=1.0\columnwidth]{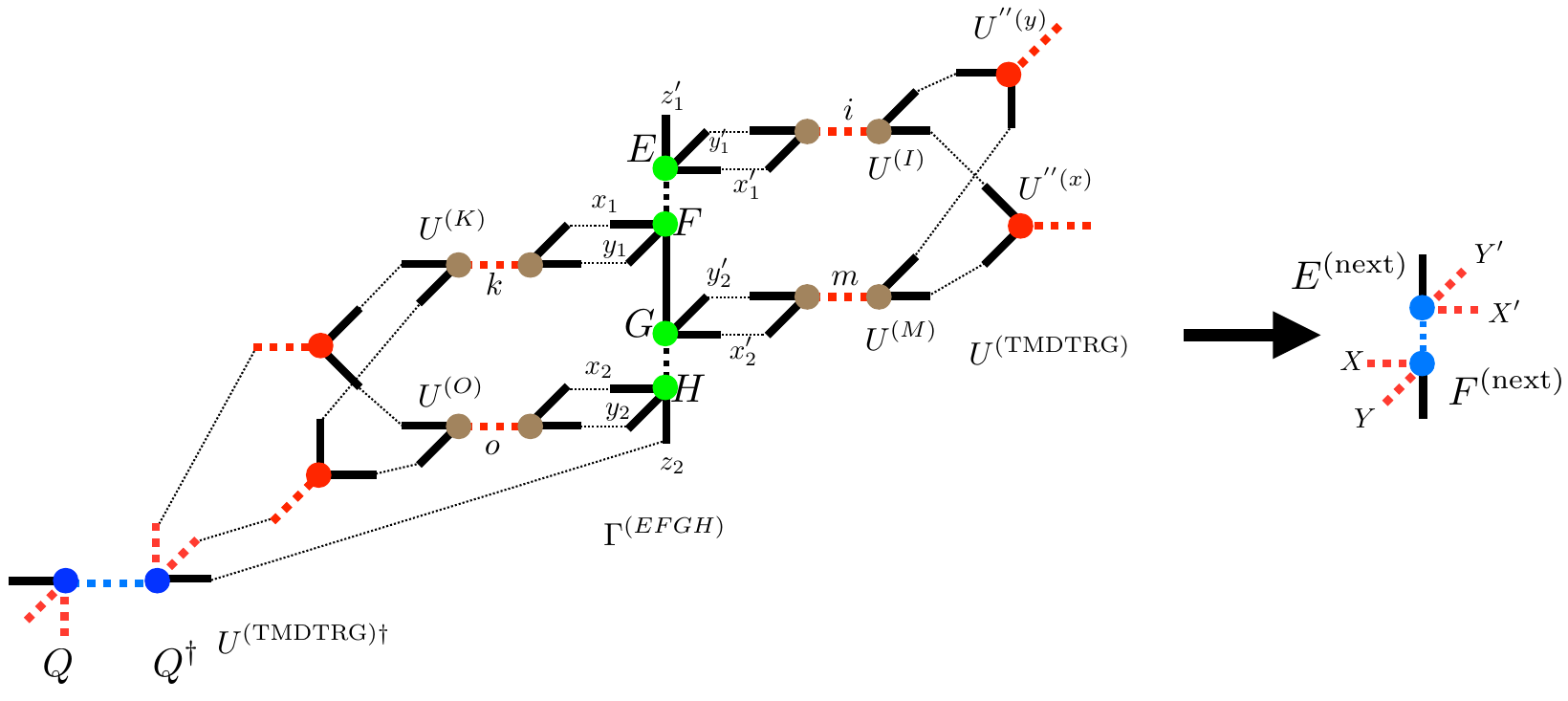}
 \includegraphics[clip,width=0.65\columnwidth]{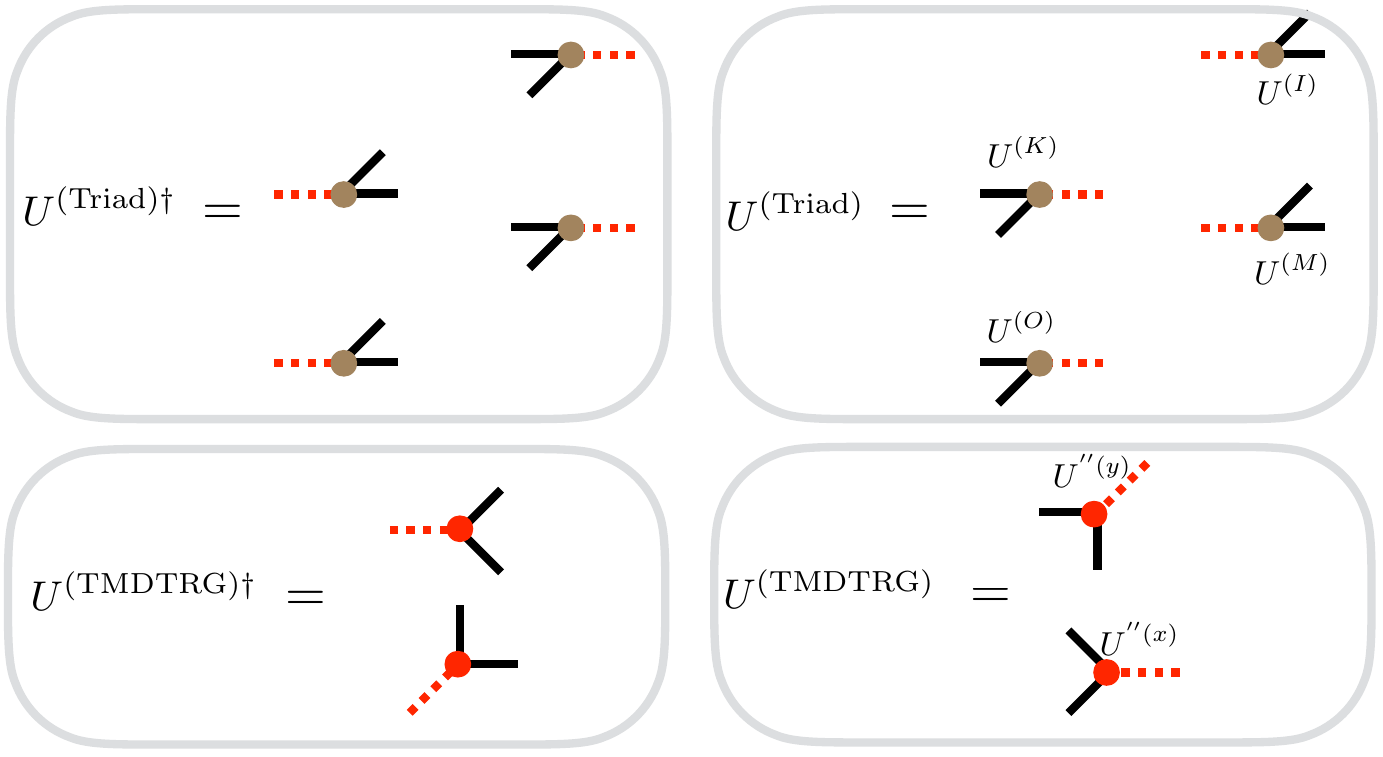}
 \caption{The contraction of the MDTRG for triad representation with the definition of each isometries $U^\mathrm{(Triad)}$ and $U^\mathrm{(TMDTRG)}$.}
 \label{Cont}
\end{figure}

We introduce additional decomposition concerning the original unit-cell tensor network for further cost reduction.
In other words, we represent all decomposition to find the triad representation using the $\Gamma^{(EFGH)}$ in the cost function. For simplicity, we consider the three-dimensional system as an example. Also, we summarize the detailed definition of each isometries in Figure \ref{Cont}.

We define the triad representation of the unit-cell tensor as the $\Gamma^{(IJKLMNOP)} = IJKLMNOP = U^\mathrm{(Triad)\dagger}\Gamma^{(EFGH)}U^\mathrm{(Triad)}$ and corresponding cost function is $|| U^\mathrm{(Triad)\dagger}\Gamma^{(EFGH)}U^\mathrm{(Triad)} - \Gamma^{(EFGH)} ||$.
The tensors $I,J,K,L,M,N,O$ and $P$ are triad tensors.
In order to find the total isometry $U^\mathrm{(Triad)}$ for triad representation, we consider the contracted tensor network $[\Gamma^{(EFGH)}\Gamma^{(EFGH)\dagger}]$ for corresponding indices, which can be identified by using Figure \ref{Cont}.
For simplicity, we calculate the isometries $U^{(I)} _{x_1 ' y_1 ' i}$, $U^{(K)} _{x_1 y_1 k}$, $U^{(M)} _{x_2 ' y_2 ' m}$, and $U^{(O)} _{x_2 y_2 o}$ and construct $U^\mathrm{(Triad)} _{abcd}= U^{(I)} _aU^{(K)} _bU^{(M)} _cU^{(O)} _d$.
Again, we note that each isometry for the triad representation also introduces the internal line oversampling parameter $r$. At any step in our calculation, we keep all indices from the randomized SVD are oversampled.

Using the triad representation of the unit-cell tensor network $\Gamma^{(IJKLMNOP)}$, we consider the triad MDTRG method.
The total isometry using $\Gamma^{(IJKLMNOP)}$ is defined as $U^\mathrm{(TMDTRG)} _{x_1x_2Xy_1y_2Y}= U^\mathrm{''(x)} _{x_1x_2X}U^\mathrm{''(y)} _{y_1y_2Y}$, where the isometries $U^\mathrm{''(x)}$ and $U^\mathrm{''(y)}$ for each direction are calculated by the $\Gamma ^{(IJKLMNOP)} \Gamma ^{(IJKLMNOP)\dagger}$.
The corresponding cost function for $Q$ is
\begin{align}
||
QQ^\dagger U^\mathrm{(TMDTRG) \dagger}\Gamma^{(IJKLMNOP)}U^\mathrm{(TMDTRG)}
-
U^\mathrm{(TMDTRG) \dagger}\Gamma^{(IJKLMNOP)} U^\mathrm{(TMDTRG)}
||.
\end{align}
We can use the matrix $Q$ to take the approximate contraction with the computational cost $O(r^3D^{d+3})$.
It also means that the course-grained tensor is 
\begin{align}
A^\mathrm{(next)}
\simeq
QQ^\dagger U^\mathrm{(TMDTRG) \dagger}U^\mathrm{(Triad) \dagger}\Gamma^{(EFGH)} U^\mathrm{(Triad)}U^\mathrm{(TMDTRG)}.
\label{TMDTRG}
\end{align}

We note that all our decomposition and the corresponding cost functions are based on the unit-cell tensor $\Gamma$, $\Gamma^{(EFGH)}$ or $\Gamma^{(IJKLMNOP)}$.
This property guarantees that all our approximations focus on the same approximation region that defines the cost function, and the region corresponds to that of the original HOTRG.

In the cost function representation, we can straightforwardly extend our method to other methods, such as the variational optimization or the projective truncation method, to find the isometry.
Especially, MDTRG with the triad representations contains independent isometries for each direction, and then how to optimize these isometries is highly non-trivial.
As we will show in this proceeding, our most straightforward choice, which defined the isometry by the simple SVD of the unit-cell tensor network, produces consistent results with the HOTRG.
We may find more sophisticated isometries by the different algorithms in future works. 

\section{Numerical calculation in three-dimensional system}

We apply our randomized HOTRG, MDTRG, and triad-MDTRG to the three-dimensional Ising model.
The partition function Z is 
\begin{align}
Z\equiv
\mathrm{Tr}\sum_{i}
A_{x_iy_iz_ix' _iy' _iz' _i}
\end{align}
where the tensor $A$ are defined $2\times 2$ matrix $W$ as following,
\begin{align}
A_{xyzx'y'z'}
=
\sum_{k=1} ^2
W_{kx}W_{ky}W_{kz}
W_{kx'}W_{ky'}W_{kz'}.
\end{align}
The $2 \times 2$ matrix $W$ are defined as 
\begin{align}
W
=
\left(
\begin{matrix}
\sqrt{\mathrm{cosh}\beta},\sqrt{\mathrm{sinh}\beta}\\
\sqrt{\mathrm{cosh}\beta},-\sqrt{\mathrm{sinh}\beta}
\end{matrix}
\right)
\end{align}
with the inverse temperature $\beta$.
Our calculations are tested with the volume $V = 2^{45}$, at the critical temperature $\beta_c^{-1} = 4.5115$ with the oversampling parameter $r = 6$ and the number of QR decomposition $q=2$ for each randomized SVD.
We calculate the partition function $Z$ and the free-energy density ${\cal{F}} \equiv -\frac{1}{\beta V}\mathrm{log}Z$.

\begin{figure}[t]
\begin{minipage}[t]{0.5\columnwidth}
\centering
 \includegraphics[width = 1.0\columnwidth]{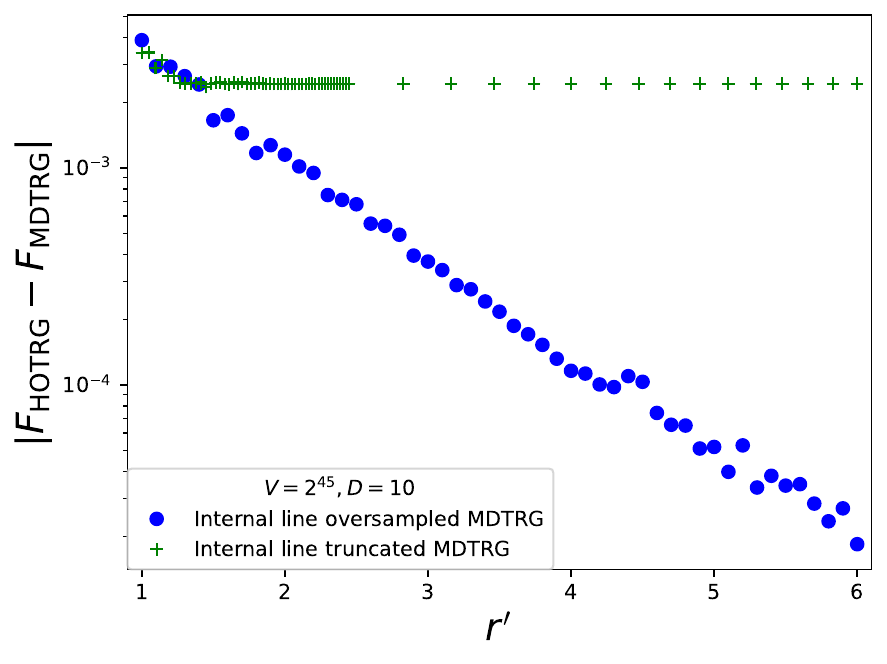}
\end{minipage}
\begin{minipage}[t]{0.5\columnwidth}
\centering
 \includegraphics[clip,width=1.0\columnwidth]{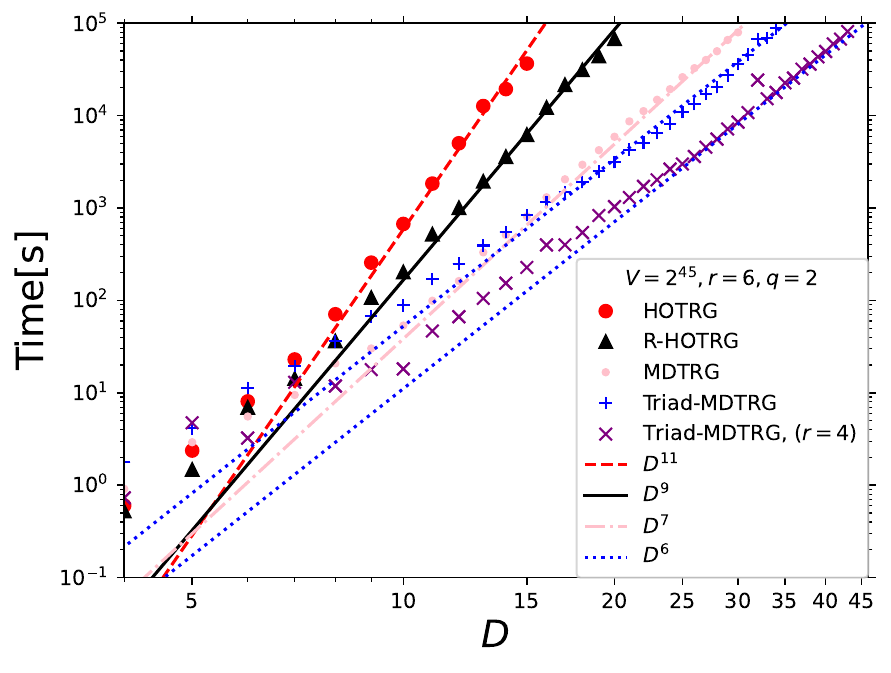}
\end{minipage}
 \caption{
 Comparison between the MDTRG with and without internal line oversampling (left) and the computational time for each method depends on the truncated bond dimension $D$ (right) for the three-dimensional Ising model at the critical temperature.}
 \label{R_dep}
\end{figure}

The left-hand side of the figure \ref{R_dep} shows the free energy dependence of the normalized oversampling parameter $r'$.
We define the normalized internal line oversampling parameter $r' = r$ for MDTRG and corresponding parameter $r' = \sqrt{r}$ without internal line oversampling MDTRG since the cost of the internal line oversampled MDTRG is $O(r^2D^{2d+1})$ and the cost of the internal line truncated MDTRG is $O(rD^{2d+1})$.
Using the normalized parameter $r'$, both methods need the order $O(r'D^{2d+1})$ cost.
We compare the precision of the MDTRG to the HOTRG, and the result shows that the internal line oversampled MDTRG converges to the correct HOTRG result depending on $r'$, while the internal line truncated MDTRG does not. The right-hand side of the figure \ref{R_dep} shows that the calculation time depends on $D$.
The result is consistent with our estimation, as described in the previous sections.

\begin{figure}[]
 \centering
 \includegraphics[clip,width=0.7\columnwidth]{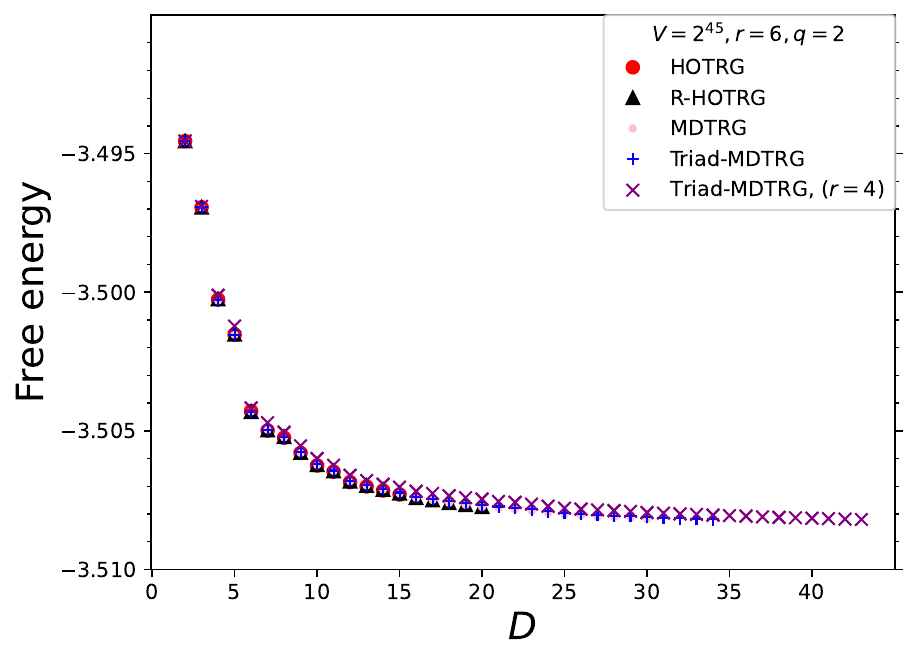}
 \caption{The free energy in the three-dimensional Ising model at a critical temperature depends on the truncated bond dimension $D$.}
 \label{3d_energy_D}
\end{figure}

Figure \ref{3d_energy_D} shows the free energy dependence of the truncated bond size $D$.
Our randomized HOTRG, MDTRG, and triad MDTRG calculations are consistent with the original HOTRG without any additional approximation except the isometry $U^\mathrm{(HOTRG)}$.
This shows that our method can only reduce the computational cost without loss of precision.

\section{Conclusion}

We show the randomized HOTRG and its further cost reduction as MDTRG and triad MDTRG.
All of our methods can be considered as cost function optimization, and this cost function representation helps us understand the approximation region and further development, including variational optimization, including projective truncation, which is not limited to the randomized SVD.

In this proceeding, we apply the randomized SVD method to the HOTRG and consider the unit-cell tensor in any cost function representations of any decompositions.
We also introduce the internal line oversampling to realize precision comparable to the original HOTRG method.
The randomized HOTRG, MDTRG, and triad MDTRG achieve the reduced cost as $O(D^{3d})$, $O(D^{2d+1})$, and $O(D^{d+3})$, respectively.
Because of the internal line oversampling and the same approximated region, our calculation also shows consistent precision compared to the HOTRG.

\end{document}